\documentstyle[prd,preprint,aps]{revtex}
\begin{document}
\draft
\tighten
\title{Structure of Pairs in Heavy Weakly-Bound  Nuclei}
\author{J.R. Bennett,$^{1,2}$ J. Engel,$^1$ and S. Pittel$^2$}
\address{ $^1$ Department of Physics and Astronomy, CB3255, University of North
Carolina, Chapel Hill, NC 27599--3255 \\
$^2$ Bartol Research Institute, University of Delaware, Newark, DE 19716}
\maketitle

\begin{abstract}
We study the structure of nucleon pairs within a simple
model consisting of a square well in three dimensions and a delta-function
residual interaction between two weakly-bound particles at the Fermi surface.
We include the continuum by enclosing the entire system in a large spherical
box.  To a good approximation, the continuum can be replaced by a small set of
optimally-determined resonance states, suggesting that in many nuclei far from
stability it may be possible to incorporate continuum effects within
traditional shell-model based approximations.
\end{abstract}

In weakly-bound nuclei far from the valley of stability, the effects of
continuum single-particle states can be significant.  The size of
these effects and the physical quantities in which they appear are largely
unexplored subjects.
Previous studies have focussed on specific issues.  Dobaczewski {\it et al.}
\cite{dob} formulated the Hartree-Fock-Bogoliubov equations in coordinate space
to study pairing near the neutron drip line.  Bertsch and Esbensen studied
pairing in light halo nuclei, using a Green function framework in which the
continuum was discretized \cite{ber}.  Otsuka, also interested in light halo
nuclei, introduced the Variational Shell Model, which implicitly includes
effects of continuum states \cite{ots}.  Our interest here is slightly
different.  Like Dobaczewski {\it et al.}, we want to understand aspects of the
structure of
``heavy" nuclei far from stability.  Our intent, however, is to study the
adequacy of relatively simple shell-model approximations to a full treatment of
the continuum, with the goal of eventually calculating beta decay in nuclei
along the r-process path.

We perform all calculations in the following simple model, similar to that used
in Ref.\ \cite{ber}:  two particles move
near the Fermi surface of an external one-body potential and interact via an
attractive delta-function two-body potential.  The one-body potential is a
finite spherical square well plus a surface (derivative) spin orbit force.  To
include the continuum simply, we enclose the entire system in a large spherical
box.

An attractive delta-function potential produces a two-particle bound state at
infinitely negative energy when diagonalized in a complete space.  To avoid the
singularity we follow the procedure of Ref.\ \cite{ber}, truncating the
continuum so that the bound state appears at zero energy, a physically sensible
value.  The procedure imposes a relation between the interaction strength and
the maximum two-particle kinetic energy; outside the well we take the
interaction strength to be 831 MeV-fm$^3$\cite{ber}, which implies a cutoff of
40 MeV.  To simulate density-dependence, we reduce the strength of the
interaction inside the well to 300 MeV-fm$^3$, a value that has proved
appropriate in shell-model studies that ignore the continuum \cite{dbl}.  We
include all partial waves up to $\ell = 7$; neither increasing this number to
$\ell =17$ nor varying the box size between 30 and 50 fm changes the results
significantly.

Although we treat only two particles as active, we would like in some measure
to
simulate the dynamics of heavy nuclei.  To do so, we adjust the parameters of
the one-body potential so that its single-particle spectrum resembles that in a
nucleus with $A \approx 90$.  The square well is therefore taken to have a
radius $R$ of roughly 5.2 fm and a depth in the range 40-50 MeV.  A
spin-orbit strength $V_{\rm s.o.}$ of about 0.4-0.5 fm$^2$ then results in the
standard
ordering of single particle levels.  We assume that all but the
last bound level, or perhaps the last few bound levels, are fully occupied and
inert.  The two active particles are then distributed over the few remaining
bound orbits and all the continuum orbits (i.e., those with positive energy) up
to the cutoff.  By changing slightly the radius and depth of the square well,
we
can alter the active single-particle levels and their energies.  Here
we focus
on states with angular momentum zero and positive parity, which span a space of
dimension $\approx 1000$ (2500) two-particle states for a box size of 30 (50)
fm.

In what follows, we compare the results of a complete diagonalization in the
space of $0^+$ states with those obtained at several levels of approximation.
We first consider
truncation of the continuum to states that are largely confined within the
square well, i.e.\ resonances.  We then further reduce the basis by
constructing
one single-particle state to represent each resonance (the resonances
have widths and comprise several such states).

First we discuss the full calculations, some representative results of which
appear in Table 1.  Here the only bound state included is the $0g_{7/2}$ level.
The energy ($E_b$) of this level ranges from about -300 keV to about -3 MeV;
the
table gives the largest components of the ground state wave function for three
choices of $E_b$.  In all three sets of calculations, the continuum
single-particle states never make up more than 25\% of the ground state
eigenvector.  Furthermore, this contribution does not depend strongly on the
energy of the active bound orbit, at least within the range specified above.
Neither of these two facts is surprising.  The continuum states that couple
most
strongly to the bound orbit through a short-range interaction are those with
large spatial overlap.  Such states typically belong to resonances and are
concentrated within the well.  Even though they make up but a small fraction of
the total continuum, the resonant states nevertheless provide nearly the full
continuum contribution to the ground state; the plethora of non-resonant
continuum states couple much more weakly, typically by factors of 1000 or more.
But the resonant states are in general well separated in energy
from the last bound state,
limiting the amount of continuum mixing. Also, as the bound-state energies
are lowered the resonances are as well, by roughly the same amount.  The energy
gap between the important states is therefore nearly independent of the Fermi
energy, and consequently so is the continuum component of the ground state.

Several other important points are demonstrated in Table 1:
\begin{itemize}
\item
For the case of a single bound $0g_{7/2}$ orbit, the dominant resonances that
admix in the ground state are the $h_{11/2}$ and the $d_{5/2}$. These are
remnants of what would be the ``next major shell" in a deeper potential well.
\item

The dominant resonance contributions are of two types.  Configurations in which
the two continuum particles occupy the same level play an important role.
There
are also, however, significant contributions in which they occupy different
levels {\it within the same resonance}.  Such admixtures are difficult
to incorporate into traditional treatments of pairing, which assume that pairs
are composed of particles in time reversed orbits.  We discuss a method for
treating the ``intra-resonance" pairing shortly.
\end{itemize}

The case reported in Table 1 has a relatively high-spin bound orbit, which
couples primarily to continuum states that also have high spins.  A somewhat
different picture emerges when the bound level has quantum numbers $s_{1/2}$
and
can strongly couple with $s$ resonances.  Even at low energies in the
continuum,
$s$ resonances are typically very broad since there is no centrifugal barrier
to
help localize them.  As a consequence, more continuum states admix into the
ground state.  Moreover, because the resonances are so broad the energy gap
separating them from the bound states is smaller and components with one bound
nucleon and one in the continuum appear.  These facts are illustrated in Table
2, which present results for a case in which the only active bound orbit is the
$2s_{1/2}$ with $E_b ~=~ -1.26~ MeV$.

In more realistic scenarios, the valence nucleons may occupy more than one
bound
orbital.
We find that very
little of import is different when
we include all the states of the bound 2s-1d-0g shell in our calculations.
The bound orbits are much more likely to be
occupied than the unbound ones, and only a few resonant states in the continuum
play a significant role.  These conclusions, which are nearly universal in this
simple model, probably do {\it not} hold, however, in extremely weakly-bound
nuclei very near the drip lines; some authors\cite{prl} have found
that the nuclear mean field may be quite different there.  In particular, shell
structure may be smoothed out so that energy gaps between single-particle
orbitals shrink.  This would no doubt cause significantly more continuum mixing
than in our calculations.  Our focus, however, is not on nuclei at the drip
lines.  We are concerned instead with nuclei such as those along the r-process
path, for which separation energies are typically 1-2 MeV \cite{rp}.
There the results of
our simple model should be relevant.  They show, in short, that  continuum
contributions to the
ground state wave function are fairly weak and are dominated by resonances.

These conclusions suggest that in treating nuclei away from stability, but
not too near the drip lines, we may be able to simply eliminate non-resonant
states without significantly changing the results. To test this idea, we need
a working definition of a resonant state. To this end we define the
quantity
\begin{equation}
\label{prob}
I  = \int_0^{\cal R } dr ~ u^2(r),
\end{equation}
where $u(r)$ is a normalized single-particle continuum radial wave function
and ${\cal R} \sim R + 3$ fm.
The resonant states are obviously those with large $I$. But any given
resonance has a width, i.e.\ is spread over several of the continuum
single-particle states. As a working definition, we will therefore count
as resonant any single-particle continuum state within one
standard deviation of the local maximum of $I$ (corresponding to the peak
of the resonance). If two resonances overlap, as sometimes occurs for $s$
waves, we include all the states in both peaks.

We may now truncate the continuum to include only these resonant states,
diagonalize, and compare the results with those from the full space.  Rather
than compare spectra, we focus on the single-particle density, since it is
likely to be particularly affected (mostly outside the nuclear radius) by
continuum contributions in weakly-bound nuclei.  In the two-particle model the
density is given by
\begin{equation}
\rho ({\bf r}) = \delta({\bf r} - {\bf r_1}) + \delta({\bf r} - {\bf r_2}).
\end{equation}
Figure 1 shows the one-body
ground-state densities resulting from a single bound $0g_{7/2}$ level at energy
$E_b = -1.24$ MeV, plus continuum admixtures.  In addition to the full density
distribution, we show the densities that arise when only the bound state is
included and when only the bound state and resonances states are included.
The ground state energies for the three calculations are $-6.33$ MeV,
$-2.48$ MeV and $-5.61$ MeV, respectively.

The main effects of the continuum {\it in the full result} are to increase the
density near the origin and outside the well (the latter effect has been
carefully noted in Refs.  \cite{dob,ber}; the irregularities near 50 fm in
Figure 1 are caused by the box).  The truncation to resonances in most regions
does not significantly change the density.  The biggest deviations occur just
outside the well, where the approximate result is too large, and in the long
tail, where it oscillates around the full density.  But for processes such as
beta decay that are not extremely sensitive to details of the radial wave
function, the truncation should serve admirably.

As noted earlier (see Table 1), the principal resonance admixtures involve
either both particles in the same single-particle state or two particles in
close-lying continuum states within a given resonance (and thus with the same
$l$ and $j$ quantum numbers).  As we also noted, the latter admixtures are
difficult to incorporate in traditional shell-model treatments of pairing,
based, e.g., on the broken-pair (generalized-seniority)
approximation\cite{sen}.
With this
in mind, we have tried to identify an optimum single-particle state
representing
each resonance, which is usually a set of several continuum levels.  There is
no obviously best procedure for doing this, but a reasonable prescription is to
define the single state that represents the set contained in a given resonance
by
\begin{equation}
\label{e:single}
|R(l,j)\rangle = N \sum_i \sqrt{\frac{I_{ilj}}{E_{ilj} - E_0}} |ilj\rangle~,
\end{equation}
where $N$ is a normalization constant, $|ilj\rangle$ are the states in
a given $(l,j)$ resonance, $E_0$ and $E_{ilj}$ are the single-particle energies
of the bound and resonance states, respectively, and $I_{ilj}$ is the quantity
defined in Eq.\ (\ref{prob}) as a measure of the probability of finding a
particle
near the potential well.  Now, however, we include the specific quantum numbers
$i,~l$ and $j$ for the continuum single-particle state, to make evident the use
of a separate state $|R(l,j)\rangle$ for each resonance with given $l$ and $j$
(the label $i$
distinguishes the different single-particle states in a given resonance).  If
there is more than one bound state, we take $E_0$
to be the average bound-state single-particle energy.  This definition of
$|R(l,j)\rangle$ follows from a) the requirement that it roughly reproduce (to
lowest
order in perturbation theory) the contribution of the set of states
$|ilj\rangle$ to
the ground-state eigenvector, and b) the assumption that the wave functions
corresponding to the states $|ilj\rangle$ (with fixed $l$ and $j$) are roughly
proportional inside the well.  Figure 2 shows the results of this approximation
when the active bound state has quantum numbers $0g_{7/2}$ and energy $E_b~=~
-1.24$ MeV, alongside the corresponding results for the full calculation and
the
truncation to resonant states discussed earlier.
[The ground state energies for the three calculations are $-5.79$ MeV,
$-6.33$ MeV and $-5.61$ MeV, respectively.]
The linear combination of
states within the resonance prescribed by Eq.\ (\ref{e:single}) naturally
incorporates the effects of pairing between different states (within a given
$l,~j$ resonance).  Little, apparently, is lost in approximating each resonance
by one well-chosen single-particle state.

This conclusion is even stronger for $J \neq 0$ pairs, since their components
are
coupled less by a short-range interaction.  The
approximation of each resonance by a single state should therefore
also work in many nucleon systems (where
$J \neq 0$ pairs can play an important role), allowing
the use of techniques familiar from the long history of the shell model.
We have not verified this explicitly since
we treat only two particles here, and it is possible that the approximation
will break down with the addition of more particles.  We
believe this to be unlikely, however.  The two-body matrix elements connecting
bound and resonant states with non-resonant continuum states are very small and
it is hard to see how the latter could ever play an important role.  We
therefore intend to follow the procedure discussed here in calculating beta
decay from nuclei along the r-process path, existing treatments of
which\cite{klap,beta,rproc} are schematic and not totally reliable.  These
nuclei, though far from stability, are nevertheless also sufficiently far from
the drip lines that the approach we have described here
should suffice.  Although more complicated techniques --- for example the
coordinate space HFB of Ref.\ \cite{dob} --- may be necessary for accurate
treatments of nuclear radii and other quantities that are sensitive to small
features in the spatial wave function, the approximations discussed above
together
with, e.g., the broken-pair approximation should be adequate for beta decay
provided one can reliably fix a single-particle potential and a two-body
interaction.

\bigskip

The authors acknowledge helpful discussions with W. Nazarewicz and H. Esbensen.
This work was supported by the U.S. Department of Energy under Grant
DE-FG05-94ER40827 and by the National Science
Foundation under Grant No. PHY-9303041.

\newpage

\centerline{\bf Figure Captions}

\bigskip
\bigskip

\noindent
{\bf Figure 1.} One-body densities for calculations involving a bound
$0g_{7/2}$ state with energy $-1.24$ MeV.
The solid line is the full density, the
long-dashed line is the density when nonresonant continuum states are omitted,
and the short-dashed line is the density when the continuum is omitted
entirely.

\bigskip
\bigskip

\noindent
{\bf Figure 2.} One-body densities for calculations involving a bound
$0g_{7/2}$ state with energy $-1.24$ MeV.
The short-dashed line is the full density, the long-dashed
line is
the density when nonresonant continuum states are omitted, and the solid line
is the density when each resonance is replaced by one optimal single-particle
state (see text).
\newpage
\bigskip
\bigskip
\centerline{\bf Table 1}
\bigskip
\bigskip
\begin{center}
\begin{tabular}{lcccc|ccccc|cccc}
\hline
\multicolumn{5}{c}{\rm $E_b$} ~=~ -2.88~MeV
& \multicolumn{5}{|c}{\rm $E_b$} ~=~ -1.24~MeV &
\multicolumn{4}{|c}{\rm $E_b$} ~=~ -0.286~MeV \\ \hline
\hline
{}~{\rm Amp.}~
& ~{$\ell_j$}~ & ~{\rm $E_1$}~~~ & ~{\rm $E_2$}~ &  &
{}~~~~{\rm Amp.}~
& ~{$\ell_j$}~ & ~{\rm $E_1$}~~~ & ~{\rm $E_2$}~ &  &
{}~~~~{\rm Amp.}~
& ~{$\ell_j$}~ & ~{\rm $E_1$}~~~ & ~{\rm $E_2$}~ \\ \hline
\hline
   0.891 & $g_{7/2}$ &  -2.88&   -2.88 & & 0.889 & $g_{7/2}$  & -1.24  & -1.24&
   &0.878 & $g_{7/2}$  &-0.286  &-0.286 \\
  -0.307 & $h_{11/2}$ &   1.58 &   1.58 & &-0.213 & $h_{11/2}$&   3.17&
3.24&
  &-0.294 & $h_{11/2}$&    4.07&    4.07 \\
   0.191 & $d_{5/2}$ &   .023 &   .023& &-0.159 & $h_{11/2}$&    3.24&    3.24&
  & 0.110 & $d_{5/2}$  &  1.72  &  2.30 \\
   -.091 & $h_{9/2}$ &   9.38 &   9.38& &-0.143 & $h_{11/2}$&    3.17&    3.17&
  & -.102 & $h_{11/2}$&    4.07&    4.37 \\
    .071 & $i_{13/2}$&    11.2&    11.9& &0.117 & $d_{5/2}$  &  1.07  &  1.43&
  &  .096 & $d_{5/2}$  &  1.72  &  1.72 \\
    .070 & $i_{13/2}$&    11.9&    11.9& &.102 & $d_{5/2}$  &  1.07  &  1.07&
  & -.075 & $h_{9/2}$ &   11.7 &   11.7 \\
   -.064 & $h_{9/2}$ &   8.64 &   9.38& &-.077 & $h_{9/2}$ &  10.2  &  11.1&
  &  .074 & $i_{13/2}$&    14.4&    14.4 \\
   -.044 & $h_{9/2}$ &   9.38 &   10.7& &.070 & $i_{13/2}$&    12.9&    13.9&
  &  .068 & $d_{5/2}$  &  1.72  &  3.20 \\
    .043 & $d_{3/2}$  &  2.49  &  3.29& &.067 & $d_{5/2}$  &  1.43  &  1.43&
  &  .066 & $d_{5/2}$  &  1.21  &  1.72 \\ \hline
\end{tabular}
\end{center}

\vspace{0.2in}

\noindent
Ground state amplitudes (Amp.) for three sets of calculations involving a
bound $0g_{7/2}$ state and its mixing with the continuum.
The energy of the bound
level is denoted $E_b$.
The levels are labeled by $j$ and $l$ values,
together with the unperturbed energies of the two particles.
All energies are in MeV; the bound state has negative energy.

\newpage
\centerline{\bf Table 2}
\bigskip
\bigskip
\begin{center}
\begin{tabular}{lccc}
\hline
\hline
{}~{\rm Amp.}~  & ~{$\ell_j$}~ & ~{\rm $E_1$}~~~ & ~{\rm $E_2$}~ \\ \hline
\hline
0.944 &  $s_{1/2}$ &  -1.26 &  -1.26 \\
0.110 & $s_{1/2}$ &  0.994 &  -1.26 \\
.103 & $s_{1/2}$ &   1.74 &  -1.26 \\
.101 & $s_{1/2}$ &  0.449 &  -1.26 \\
-.092 & $h_{9/2}$ &   4.16 &   4.16 \\
.089 & $s_{1/2}$ &   2.69 &  -1.26 \\
.087 & $i_{13/2}$ &   6.53 &   6.53 \\
.075 & $s_{1/2}$  &  3.84 &  -1.26 \\
.064 & $s_{1/2}$  & 0.114 &  -1.26 \\
.062 & $s_{1/2}$  &  5.18 &  -1.26 \\ \hline
\end{tabular}
\end{center}

\vspace{0.2in}

\noindent
Ground state amplitudes involving a bound $3s_{1/2}$ state, with
${E_b}~=~-1.26~$ MeV. All other notation is as in Table 1.

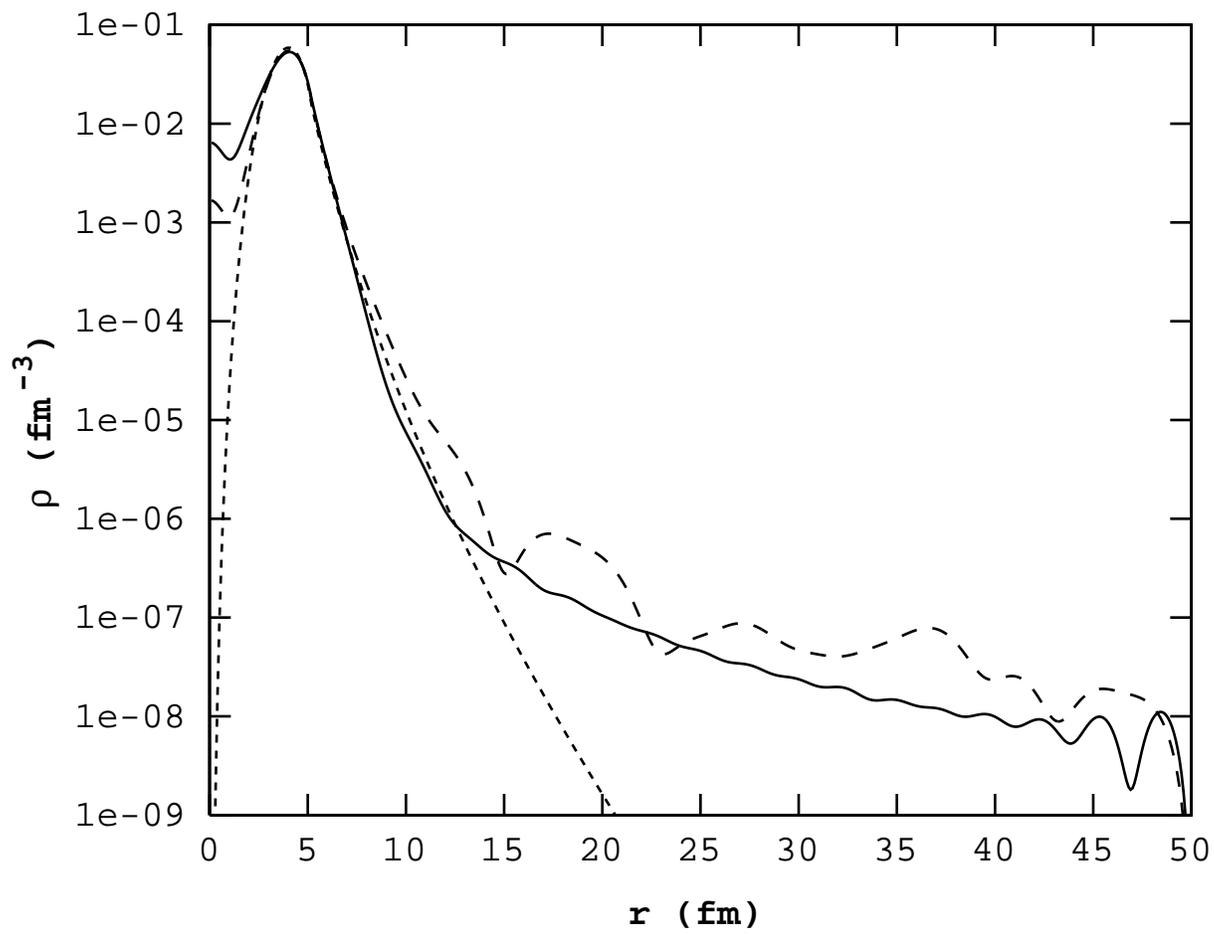
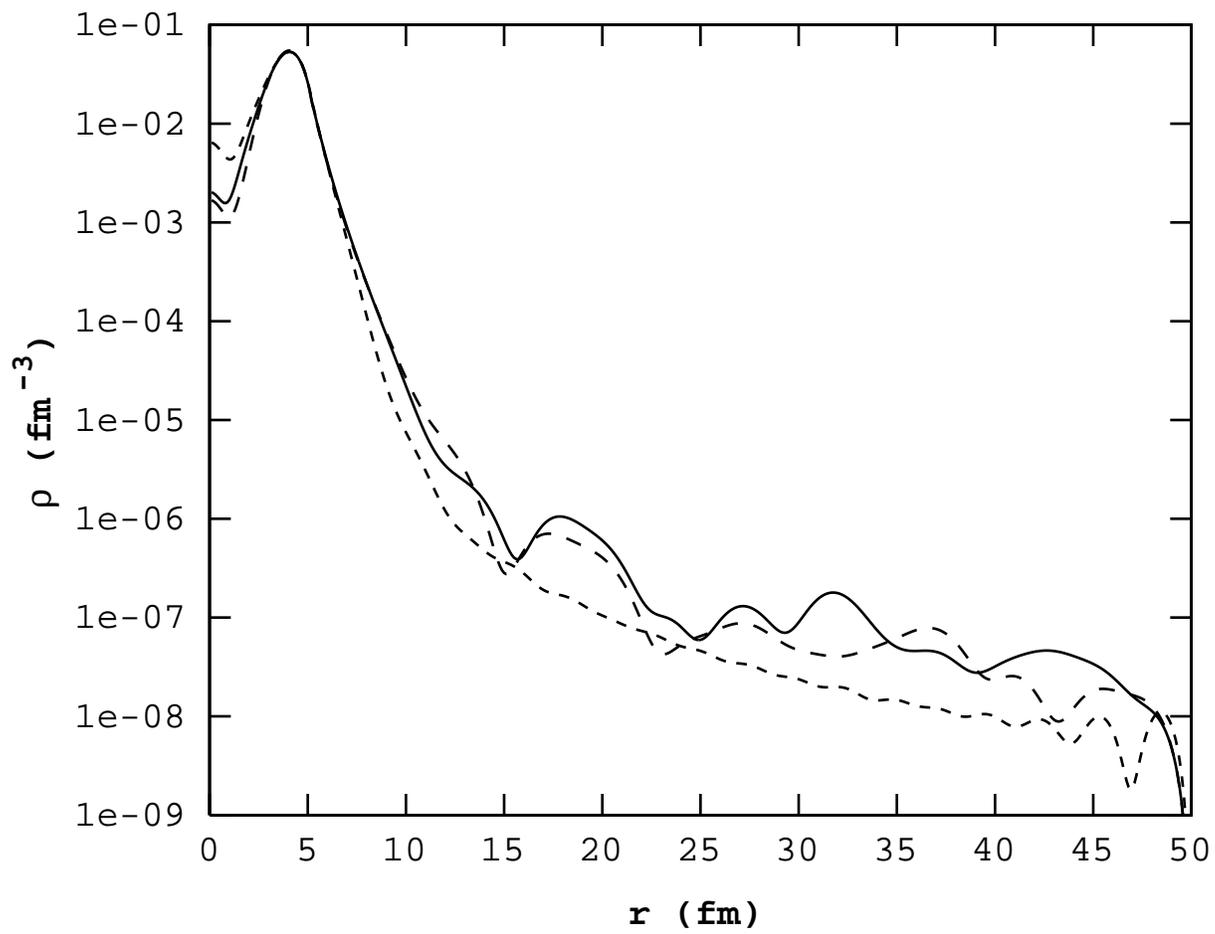
\end{document}